\documentclass{article}

\usepackage[utf8]{inputenx} 
\usepackage[T1]{fontenc}    

\usepackage{lmodern}           
\usepackage[scaled]{beramono}  
\usepackage[final]{microtype}  

\usepackage{arxiv}

\usepackage{url}            
\usepackage{booktabs}       
\usepackage{amsfonts}       
\usepackage{nicefrac}       
\usepackage{microtype}      
\usepackage{lipsum}		
\usepackage{graphicx}
\usepackage{natbib}
\usepackage{doi}
\usepackage{todonotes}
\usepackage{amsmath}

\DeclareRobustCommand{\uvec}[1]{{%
		\ifcsname uvec#1\endcsname
		\csname uvec#1\endcsname
		\else
		\bm{\hat{\mathbf{#1}}}%
		\fi
}}

\newcommand\blfootnote[1]{%
  \begingroup
  \renewcommand\thefootnote{}\footnote{#1}%
  \addtocounter{footnote}{-1}%
  \endgroup
}

\usepackage{graphicx} 

\title{Direct in-situ observations of wave-induced floe collisions in the deeper Marginal Ice Zone}

\author{
Lars Willas Dreyer$^*$ $^+$ \\
University of Oslo \\
Department of Mathematics \\
Section for Mechanics \\
\texttt{larswd@math.uio.no}
\And
Jean Rabault$^*$ \\
Norwegian Meteorological Institute \\
IT Department \\
\texttt{jean.rblt@gmail.com}
\And
Atle Jensen \\
University of Oslo \\
Department of Mathematics \\
Section for Mechanics \\
\texttt{atlej@math.uio.no}
\And
Ingrid Dæhlen \\
University of Oslo \\
Department of Mathematics \\
Section for Statistics and Data Science \\
\texttt{ingrdae@math.uio.no}
\And
Yngve Kristoffersen \\
University of Bergen\\
Department of Earth Sciences\\
\texttt{Yngve.Kristoffersen@uib.no}
\And
Øyvind Breivik \\
Norwegian Meteorological Institute \\
R\&D Department \& \\
University of Bergen \\
Geophysical Institute \\
\texttt{oyvindb@met.no}
\And
Gaute Hope \\
Norwegian Meteorological Institute \\
R\&D Department \\
\texttt{gauteh@met.no}
}

\date{January 2025}

\begin{document}

\maketitle

\begin{abstract}
    Ocean waves propagating through the Marginal Ice Zone (MIZ) and the pack ice are strongly attenuated. This attenuation is critical for protecting sea ice from energetic wave events that could otherwise lead to sea ice break-up and dislocation over large areas. Despite the importance of waves-in-ice attenuation, the exact physical mechanisms involved, and their relative importance, are still uncertain. Here we present direct in situ measurements of  floe-floe interactions under the influence of waves, including collisions between adjacent floes. The collision events we report are aligned with the incoming wave direction, and phase-locked to the wave signal, which indicates that the individual collisions we detect are wave-induced. The observations indicate a possible correlation between sudden increases in wave activity and the frequency of floe-floe collisions.
\end{abstract}

\blfootnote{$^*$: both authors contributed equally to this work.}
\blfootnote{$^+$: corresponding author.}

\section{Introduction}
\label{eq:intro}

The polar sea ice is an important regulator of the global climate \citep{budikova2009role}. In particular, polar sea ice limits ocean-air fluxes of heat \citep{wettlaufer1991heat,ivanov2019arctic} and gases \citep{soren2011sea}, limits fetch and wave growth \citep{thomson2022wave}, and the albedo of the ice sheets is a well known factor limiting the solar energy absorption in the polar regions \cite{bader2011review}.

The outermost zone of the polar ice sheets is called the Marginal Ice Zone (MIZ) and is defined as  the area along the ice pack that is affected by open ocean processes \citet{wadhams1986effect,barber2015selected}. In the MIZ, ocean waves are strongly attenuated, preventing break-up of the ice further in \citep{kohout2008elastic,bennetts2012model}. The details of the mechanisms causing wave-in-ice attenuation are still unclear, and a number of processes have been suggested in the literature, including wave refraction and diffraction by individual floes \citep{bennetts2010three}, viscous damping at the ice-water interface \citep{zhao2015modeling,sutherland2016observations,sutherland2019two}, viscoelasticity \citep{zhao2018three,zhang2021theoretical}, turbulence production and dissipation \citep{smith2019ocean}, and floe-floe interactions \citep{loken2022experiments,herman2018wave}. Uncertainties related to waves-in-ice damping mechanisms and intensity are currently limiting the accuracy of coupled numerical weather prediction and climate models of the polar regions, and significant efforts are now going into improving their accuracy \citep{boutin2020towards}. This uncertainty stems both from the large spread in different waves-in-ice attenuation parametrizations and the significant discrepancies with in-situ observations \citep{voermans2021wave}.

Collisions between ice floes is believed to be an important type of floe-floe interaction, yet their relative importance as an attenuation mechanism, and the exact nature on how these collisions attenuate waves has been a topic of discussion in the academic community. These collisions were first observed in a series of field measurements in the Bering strait and off the coast of Greenland, see \citet{martin1987high} and \citet{martin1988ice}, but the phenomenon is seldom discussed in the literature. The literature on collisions was summarized in section 2.2 in \citet{squire1995ocean}, while some mechanisms for wave attenuation was further summarized in the review articles  \citet{squire2007ocean} and \citet{squire2020ocean}, although the latter two of these reviews does not discuss floe-floe collisions in detail. In particular, \citet{li2018laboratory} observed in laboratory experiments that collisions can take place and induce significant wave energy dissipation. \citet{smith2020pancake}  observed collisions between ice pancakes in the newly formed MIZ area. \citet{rabault2019experiments}  used optical methods to observe collisions between ice chunks and the associated water pumping in a small scale laboratory experiment. This demonstrated that floe-floe collisions can cause wave-in-ice energy dissipation due to both the inelastic energy losses in the collisions themselves, as well as the increased water turbulence levels and effective eddy viscosity induced in the water immediately underneath the ice. Following these results, the study by \citet{loken2022experiments} found in idealized field experiments that floe-floe collisions can amount to up to 45 \% of the energy dissipated when ice floes move close to each other, and confirmed that collisions, when occurring, are at the origin of complex water jets and eddies. Furthermore, floe-floe collisions are also believed to be strongly related to the floe size distribution and likely to play a role in the merging and breaking of ice floes \citep{shen1991one,herman2018wave}. These collisions are believed to attenuate incoming waves both through the dissipation of energy from the collisions themselves and through increased eddy viscosity resulting from the injected turbulent kinetic energy. The dissipation due to floe-floe collisions has also been modeled by \citet{shen1998wave}, whose model found floe-floe collisions to be the predominant wave attenuation mechanism in compact pancake floe fields, and \citet{rottier1992floe}, who developed one of the first models on floe collision rates. While the early observations by \citet{martin1987high} and \cite{martin1988ice}, are important, the MIZ and polar regions are undergoing rapid changes due to anthropogenic climate change \citep[chapter 5.1]{kinda2013monitoring}. The sea ice is getting thinner  (going from a mean of 3.64 m to 1.89 m from 1980 to 2008) \citep{kwok2009decline}, younger  \citep{maslanik2007younger}, and the MIZ constitutes an increasingly large part of the total polar ice sheets \citep{strong2013arctic}. Hence, continued monitoring of the arctic ice sheets are important to adequately quantify if and how the dynamics in the area are changing with the climate.  
 Yet, the amount of observations of the phenomenon of floe-floe collisions has been low in the recent years.\citet{noyce2023identification}  recently reported time series of floe-floe collisions in the outer Antarctic MIZ during a polar cyclone. Recently, \citet{rabault24collisions} has presented observations that may indicate that sea ice convergence could modulate the intensity of floe-floe collisions, with large impact on wave in ice attenuation.

Here, we provide data showing collisions between ice floes in the arctic MIZ, and discuss the causal mechanisms and circumstances surrounding these events. We find more than 60 acceleration residuals in the horizontal plane that lie outside the 99.9\% probability region in two time series. This corresponds to more than five times the expected amount of extreme events, and is analogous to a $3\sigma$-event in the one dimensional setting. These collisions are aligned with the dominating wave direction, and phase locked with the wave signal, strongly suggesting that the individual collisions we detect are wave-induced.  

In the following, we start by presenting our observational dataset. We then describe the methodology used to detect collisions in our acceleration time series, and to confirm that these are related to the incoming waves. Finally, we discuss our findings and their implication for wave in ice damping, and we suggest additions to the firmware of existing wave in ice buoys to allow the retrieval of large datasets of in-situ floe-floe collision occurrences.

\section{Data and Methodology}
\label{sec:data}

\subsection{Buoy deployment and raw data recovery}
\label{sec:expedition}

A total of 6 waves in ice buoys (OpenMetBuoy-v2018, OMB-v2018 for brevity, see \citealt{rabault2020open}) were deployed during the summer 2020 on the Yermak Plateau in the MIZ north of Svalbard from the Research Hovercraft R/H Sabvabaa \citep{hall2009r}. The use of a hovercraft allowed navigation of the MIZ in a variety of conditions, with a much lower environmental footprint than a traditional icebreaker.

The OMB-v2018 performs wave measurements using a Vectornav VN100 Inertial Motion Unit (IMU), and uses GPS to determine its position. The wave motion data are sampled over 20-minute intervals every four hours from records of vertical acceleration sampled at 10 Hz by the VN100 \citep{rabault2017measurements}. The VN100 has built-in accelerometer, gyroscope, and magnetometer, and performs on-board Kalman filtering \citep{rabault2016measurements,rabault2020open}. This filtering ensures that the full 3-dimensional acceleration experienced by the buoy in the North-East-Down (NED) frame of reference is captured in the time series. Due to the presence of large batteries and cables carrying electrical currents in close vicinity to the VN100, the magnetometer data come with significant uncertainty and orientation relative to north should only be considered as an indication. The accuracy of the VN100 IMU is $5\cdot 10^{-3}$ g, with g being the acceleration of gravity. A comprehensive list over the tests and validations done on the VN100 IMU on the OMB performed prior to deployment is given in section 2.1 in \citet{rabault2020open}.

The OMB-v2018 performs in-situ data processing, computes the wave spectrum using the Welch method and adequate lowpass filtering \citep{rabault2020open}, and transmits compressed wave spectra back over the iridium network. The 6 OMB-v2018 operated over a two months period. The reader curious of more technical details is referred to \cite{rabault2023dataset} for a full description of the methodology and access to the transmitted data (which are available openly on GitHub\footnote{\url{https://github.com/jerabaul29/data_release_waves_in_ice_2018_2021/tree/main/Data/2020_July_Yamal}; note that the name "Yamal" in the URL is incorrect, actually this is on the Yermak plateau; since this URL is used as an archive, it is not modified.} ). No information about floe-floe collisions is available from the compressed spectra transmitted over satellite, and these data will, therefore, not be discussed in the following article.

\begin{figure}[h!]
  \centering
  \includegraphics[width=0.9\textwidth]{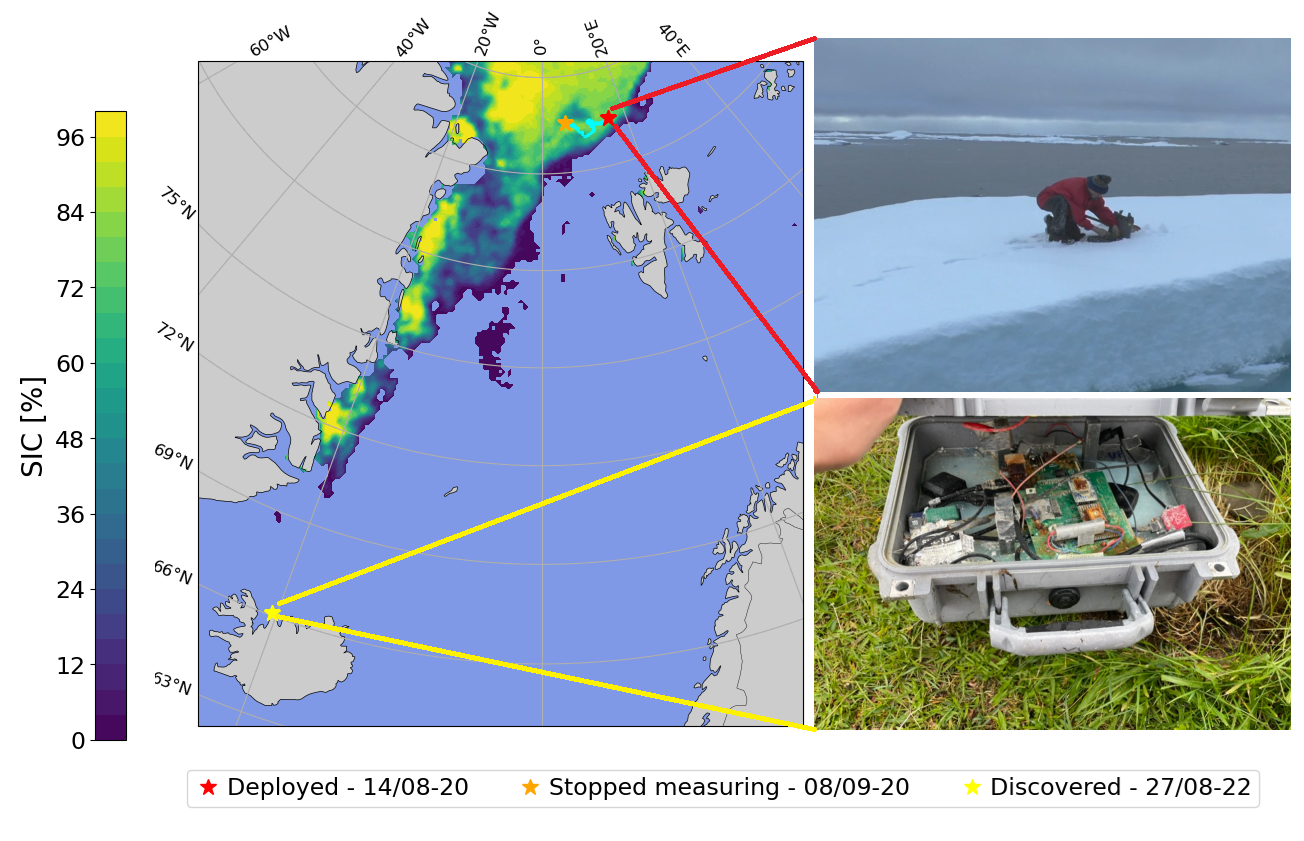}
  \caption{(A): trajectory of the recovered OMB-v2018. The buoy recovered was deployed on 2020-08-17, and stopped transmitting on 2020-09-09. Two years after deployment, the buoy was found close to the farm Hafnir, in the Skagi area in Northern Iceland. We indicate the area where the sea ice concentration (SIC) is greater than 0.05 on 2020-08-29T12Z according to the ASMR2 Sea Ice model \citep{spreen2008sea}. The buoy trajectory is shown in light blue. (B): illustration of the deployment of the buoy. (C): picture of the recovered buoy (Photo credits: Lísabet Guðmundsdóttir). Despite the seawater damage to the electronics, the SD card was intact and raw timeseries data could be recovered, which constitutes the dataset used in the present work.}\label{fig:story_map}
\end{figure}

One of the six buoys was subsequently found by an icelandic archeologist over 2 years after deployment, stranded in Northern Iceland close to Skagi, long after it had stopped transmitting data, see Fig. \ref{fig:story_map}. Following the internal design and working of the OMB-v2018 \citep{rabault2020open}, the full time series from the GPS and IMU present on-board are available on internal SD cards included in the buoy. Since the SD card is a robust device embedded in plastic and with gold plated, non corrodible connectors, all data could be successfully recovered. As a consequence, the full time series of the wave motion encountered by the buoy during its operation are available, providing a total of 142 individual files, containing each a 20 minutes IMU and GPS time series.

\begin{figure}[h!]
  \centering
  \includegraphics[width=0.9\textwidth]{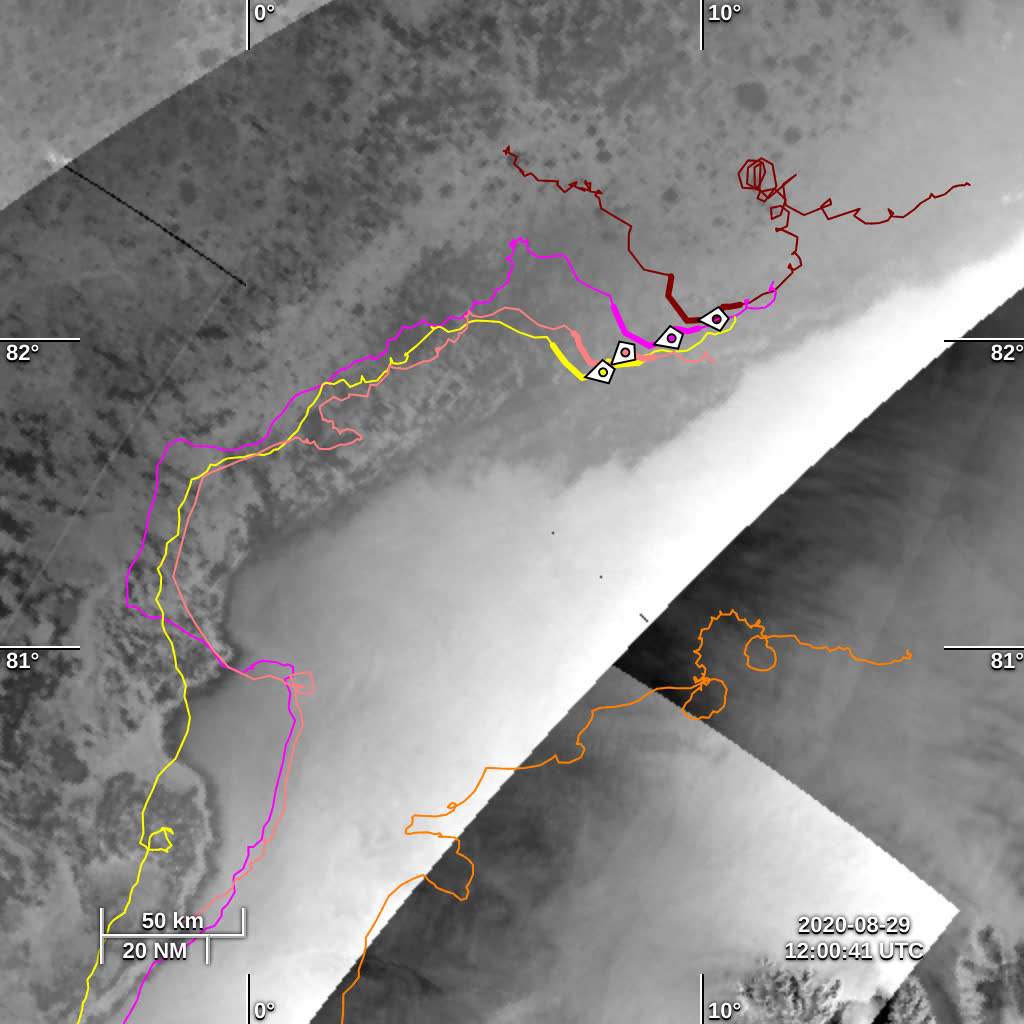}
  \caption{Trajectories of the buoys deployed 2020-08 over the Yermak plateau area, superimposed on synthetic aperture radar (SAR) images of the MIZ on 2020-08-29. The buoy that was recovered is the northernmost one, which is shown in red. The markers indicate the position of each buoy on the 2020-08-29 at 12 UTC. The figure is created using the Ocean Visual Laboratory tool (permalink: \url{https://odl.bzh/vEFWnkH0}). Sea ice is shown in gray, with dark gray spots being larger ice floes, in the SAR images, and the buoy recovered is deep in the MIZ.}
  \label{fig:OVL1}
\end{figure}

Upon analysis, the buoy that got stranded was found to be the instrument 18954 from \cite{rabault2023dataset}. The trajectory of all the buoys in this dataset, superimposed on synthetic aperture radar (SAR) images taken on the 2020-08-29, are presented in Fig. \ref{fig:OVL1}.

\subsection{Signal processing of recovered in-situ time series}

We re-compute all wave quantities from the raw data on the SD cards, including the double-integrated wave elevation, significant wave height and peak period. These are computed using the same method as \cite{kohout2015device}. Writing $\mathcal{F}$ and $\mathcal{F}^{-1}$ for the Fourier transform and the inverse Fourier transform, respectively, the surface elevation is computed as

\begin{equation}
    \eta = \mathcal{F}^{-1}\left[R(f)\mathcal{F}[a_z(t)](f) \right](t),
    \label{eq:etadef}
\end{equation}

\noindent where $R(f)$ are the same response weights used by \cite{kohout2015device}, which are derived in \cite{tucker2001waves} as:

\begin{equation}
    R(f) =
    \begin{cases}\!
    \begin{aligned}[b]
    0, & \text{ for } f < f_1, \\
     -\frac{1}{4\pi f^2}\left[1 - \cos\left(\frac{\pi(f-f_1)}{f_2-f_1}\right)\right], &   \text{ for } f_1 \leq f \leq f_2, \\
    -\frac{1}{2\pi f^2}, & \text{ for }  f_2 < f < f_c.
    \end{aligned}
    \end{cases}
    \label{eg:Rweights}
\end{equation}

In the expression above, $f_c$ denotes the Nyquist frequency, while $f_1$ and $f_2$ are threshold frequencies which we set to 0.2 and 0.3 Hz respectively, the same values used by \cite{rabault2023dataset}. This methodology performs both double integration in time and lowpass filtering, while alleviates low frequency integration noise issues common with IMU wave data.

We compute the significant wave height $H_s$ and peak period $T_p$ from the surface elevation as:

\begin{align}
    H_s &= 4\sigma_\eta = 4 \sqrt{\sum_{i}(\eta(t_i))^2}, \label{eq:Hsdef} \\
    T_p &= \frac{1}{f_\text{max}}, \label{eq:Tpdef}
\end{align}

\noindent where $f_\text{max}$ is the frequency at which the PSD reaches its maximum value. Furthermore, we estimated the steepness of the wave field, we assumed for simplicity the dispersion relation to be that of deep-water waves (which holds true to leading order, see equations 11 and 12 in \citet{meylan2018dispersion}). Hence, using the significant waveheight as the wave amplitude, and peak period as the wave period, we estimate the steepness to be: 
\begin{equation}
    a_k = \frac{4\pi^2 H_s}{gT_p^2}
\end{equation}

We also compute the wave power spectral density (PSD) of the surface elevation directly from using the Welch transform of the wave elevation $\eta(t)$ with a window size of 1024 points and an overlap of half the window size.

\begin{figure}
  \centering
  \includegraphics[width=0.9\textwidth]{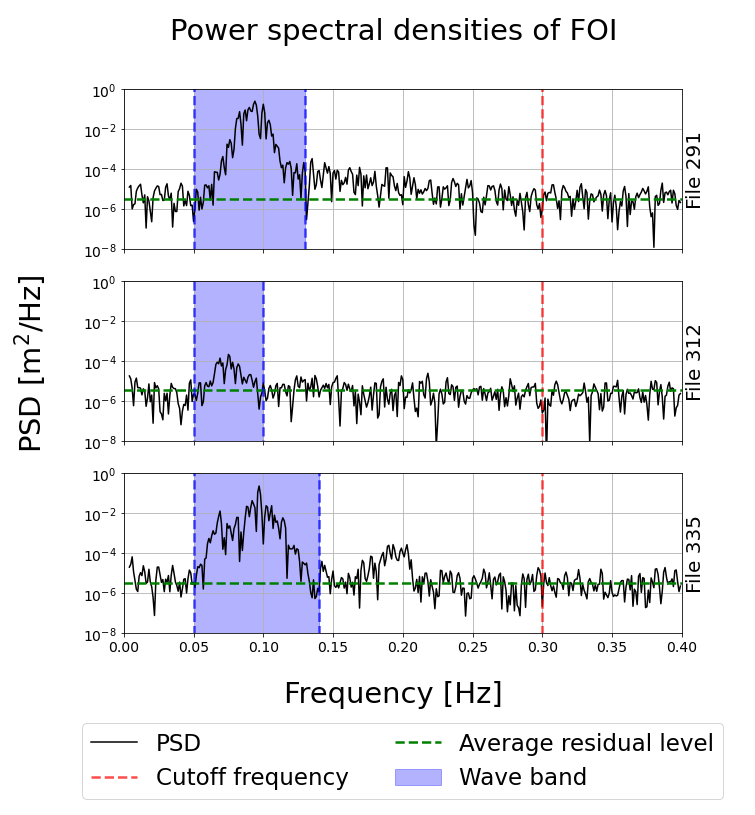}
  \caption{Illustration of the power spectral density (PSD) of the wave elevation $\eta(t)$ obtained for the two files of interest (FOIs) and File 335, which contains no confirmed collisions. The high frequency cutoff used to separate between the wave signal and the residual signal is well above the frequency range in which wave activity is observed. Clear low frequency swell, well above the instrument background noise level, is observed in all time series. Here, the cutoff frequency used for the high pass filter is shown by the red dashed line, the mean noise level by the horizontal green dashed line, and the PSD by the black line. The frequency band for the swell is highlighted with a blue background. }
  \label{fig:PSD}
\end{figure}

In our analysis, we need to separate the "wave" signal from the "non-wave" (residual) signal, which includes both sensor and signal processing noise, and (as we will demonstrate) collisions. For computing this residual, a highpass filter of the surface elevation $\eta(t)$ is implemented using a fifth order Butterworth filter (\cite{2020SciPy-NMeth}) with a cutoff frequency of $0.3$ Hz. This cutoff frequency was decided by direct observation of the wave power spectral density (PSD), so as to exclude any wave signal content from the residual. This is illustrated, for a few specially interesting time series, in Fig. \ref{fig:PSD}. A similar methodology, using the same filter, is used to also separate the wave and the residual signals on the acceleration time series.

We also estimate the wave signal phase in order to analyze the physics of the residuals signal. To do so, we compute the phase of the filtered wave signal (obtained by subtracting the residual signal from the total elevation $\eta(t)$) using the Hilbert transform \cite{2020SciPy-NMeth}, from which we obtain both the wave signal envelope and the complex wave signal phase.

\subsection{Detection of Collisions}
\label{sec:algorithm}

The measured accelerations contain three contributions: (i) wave-induced motion of the ice floe, (ii) other physically meaningful signals, in particular, as we will demonstrate in the following, floe-floe collisions, (iii) sensor and signal processing (in particular Kalman filter) noise. The latter two, i.e. (ii) and (iii), correspond to the residual signal computed above.

The noise contribution (iii) arises from the noise in the micro-electromechanical systems (MEMS) sensors used in the IMU, and its propagation through the embedded Kalman filter run on-the-fly. According to the datasheet of the IMU, and in agreement with previous observations \citep{rabault2020open}, this generally follows a normal distribution. Hence, statistically significant deviations from a normal distribution in the residuals is a likely signature of acceleration events which can not be attributed to noise. This is tested using a Pearson $\chi^2$-test. Furthermore, we compute the 99\% and 99.9\% probability areas for the residuals under an assumption that the residuals are normally distributed. Using a visual inspection of the flagged files, we can check  if an unexpectedly large amount of residuals fall outside these areas, and if so,   the residuals are most likely not normally distributed and likely contain physically meaningful signals (ii).

To test this formally, we use a Pearson chi squared test on each of the time series, testing if the residuals could
be treated as Gaussian noise against the alternative hypothesis that this is not the case
(See chapter 13 of \cite{devore2021modern} for more technical details). 
To do this, we bin the data into 10 domains. These are constructed so that, under the null
hypothesis, all bins have equal probability. This ensures that roughly the same amount of
data points should land in each bin if normally distributed. Let $I_j$ denote the $j$-th
bin, then the Pearson chi squared test statistic takes the form $K = \sum_{j=1}^{10}(N_j-E_j)^2/E_j$,
where $E_j=\{ \#\textup{Datapoints in the time series}/10 \}$ is the number of data points
we would expect landing in cell $I_j$ and $N_j$ is the number of residual landing in
bin $I_j$. Under the null hypothesis, $K$ is $\chi^2$-distributed with six degrees of freedom as we need to estimate the covariance matrix by the empirical covariance in the data. Hence, the null hypothesis is rejected on an $\alpha$ significance level if $K$
exceeds the $1-\alpha$ quantile in the $\chi^2$-distribution with six degrees of freedom. 

We test for collisions in all 142 time series recovered from the SD card.
Hence, we are performing 142 hypothesis tests simultaneously. Because of the nature of hypothesis
tests, this will lead to multiple correct null hypotheses being rejected. If we for
instance use a 1\% significance level for the tests, we would expect about seven of the 
null hypotheses being falsely rejected. To mitigate this, we performed a Bonferroni 
correction. This is done by testing each hypothesis on a $\alpha/142$-level
rather than $\alpha$-level for some fixed $\alpha$ (e.g.~1\%). This method ensures that 
the probability of the test procedure rejecting at least one correct null hypothesis is controlled,
rather than the significance level of each individual test (see \cite[p. 686]{hastie2009elements}). Hence, after performing a Bonferroni correction, the probability of the test procedure 
falsely rejecting at least one true null hypotheses is bounded by $\alpha$. 

\subsection{Ocean and sea ice model data}

Model data leveraging assimilation of satellite observations are used to estimate ocean and sea ice conditions in the area where the recovered buoy was active. In particular, we use the WAM-4km model data, run operationally at the Norwegian Meteorological Institute (MetNo), to estimate the open ocean wave conditions namely swell, wind speed and wind sea waves (see \citep{group1988wam,gunther1992wam,ardhuin2010semiempirical, breivik2022}). The corresponding model data are openly available\footnote{\url{https://thredds.met.no/thredds/catalog/fou-hi/mywavewam4archive/catalog.html}}.

Estimates for sea ice concentration and sea ice thickness were gathered from the Nordic4 model ( see \citep{budgell2005numerical,spreen2008sea}), run operationally at MetNo and openly available\footnote{\url{https://thredds.met.no/thredds/catalog/fou-hi/nordic4km-zdepths1h/catalog.html}}.

\section{Results}
\label{sec:Results}

\subsection{Collisions detection: direction and phase locking with incoming waves}

\begin{figure}
  \centering
  \includegraphics[width=0.95\textwidth]{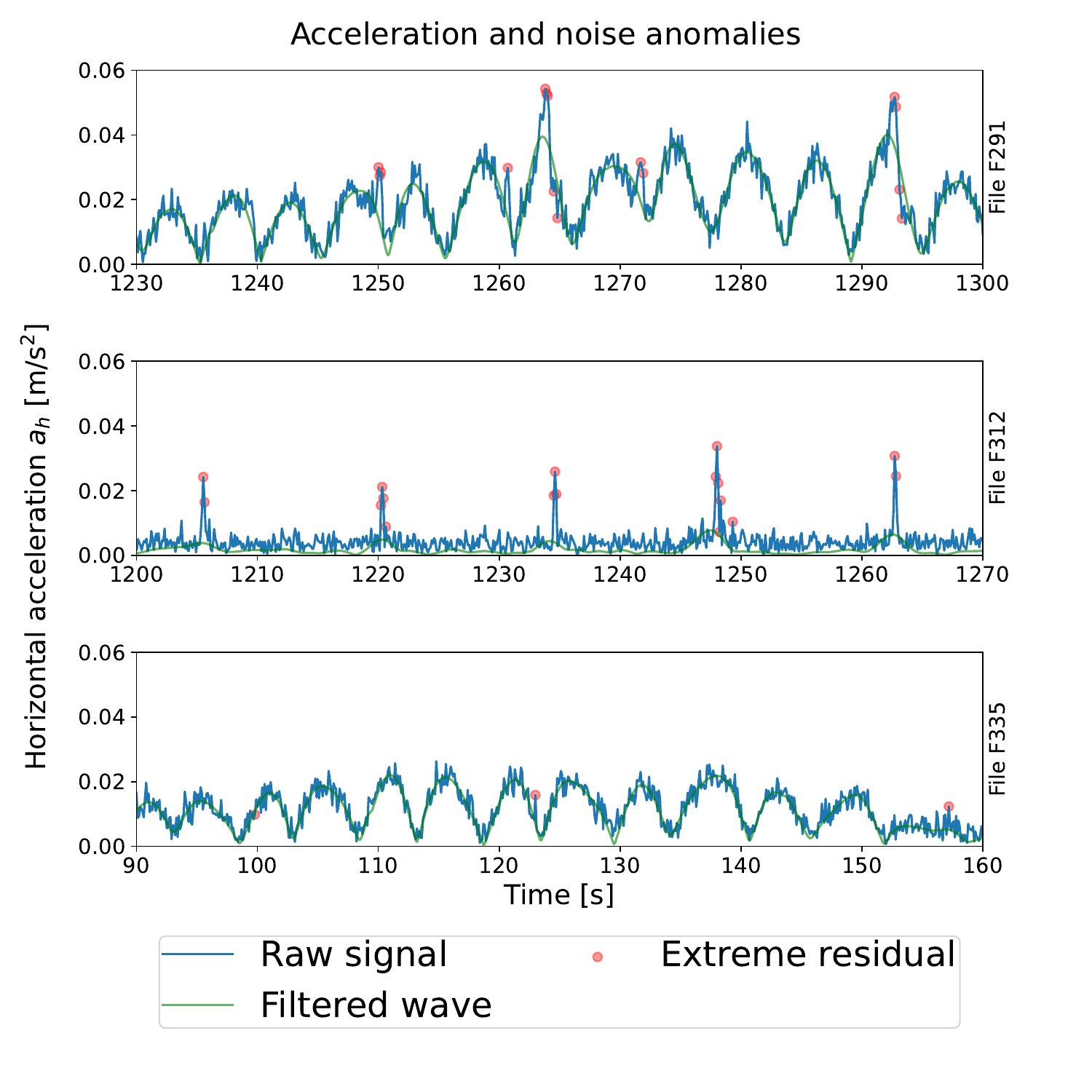}
  \caption{Illustration of the detected extreme residuals: time series from the FOIs F291 (top) and F312 (middle). A file for which the test is negative is also presented for reference (F335, bottom). The magnitude (so as to be independent of direction) of the measured acceleration vector in the horizontal plane is shown in blue. The red dots show individual extreme measurement events where the residual is outside of the 99.9\% probability region. As visible in Fig. \ref{fig:PSD}, significantly more wave energy is present in F291 than in F312. As a consequence, the horizontal wave acceleration signal is clearly visible in F291, while it is harder to see due to the much lower signal to noise ratio in the case of F312 (though Fig. \ref{fig:PSD} proves that waves are present also in F312). 
  }
  \label{fig:detectedcollisions}
\end{figure}

Using the methodology presented earlier, we find significant (according to the Pearson Chi squared test with Bonferroni correction and a 99\% confidence threshold) extreme residual events in four time series. Upon manual inspection, we found clear signs of collisions in two of these, namely files F291 and F312 (which we will refer to as the files of interest, FOIs, in the following). These events correspond to the occurrence of many extreme residual values in the VN100 acceleration time series, each lying outside the 99.9\% probability area. An illustration of the corresponding events is shown in Fig. \ref{fig:detectedcollisions}, which confirms that clear extreme events signal standing out from the wave signal and the background noise is observed. We also present the time series obtained from file F335, in which the 99\% confidence interval test is not conclusive. As visible there, though our test is negative, a few points still stand out and could be possible evidence of complex dynamics. This may indicate that our test is actually quite a bit more restrictive than it needs to be, and that more events may be present than what our test reports, but it also gives us confidence that the events flagged by our test are very likely to be more than random chance. 

\begin{figure}
\centering\includegraphics[width=0.99\textwidth]{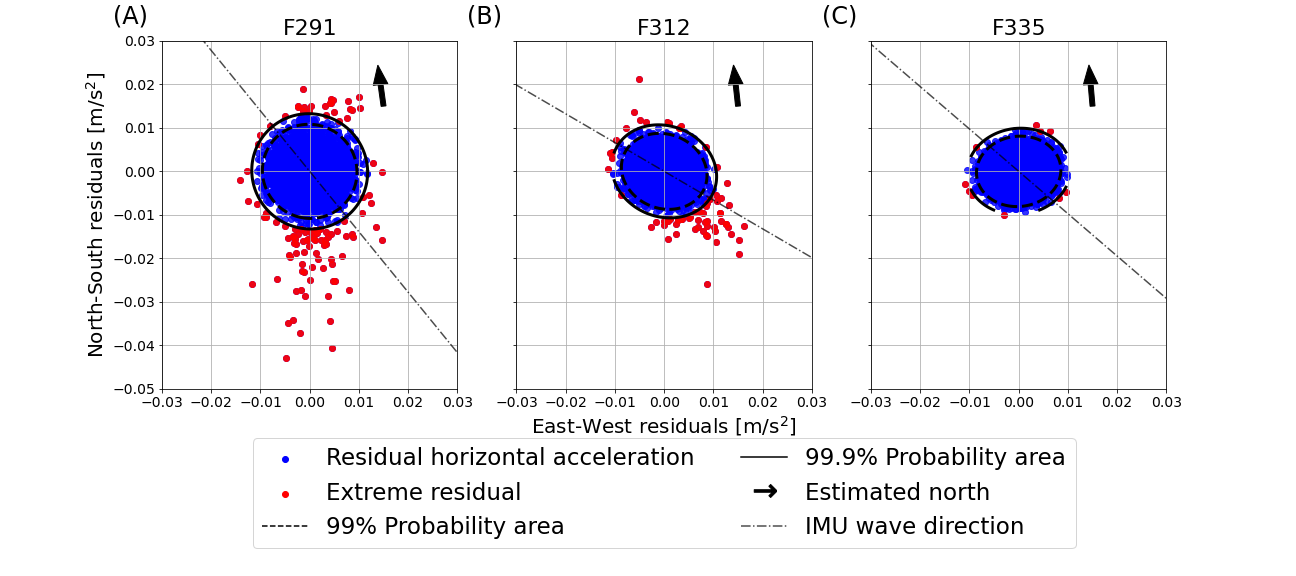}
  \caption{Illustration of the horizontal component of the residual signal in three different time series. In the FOIs (file 291 (A) and 312 (B)), there are clear extreme values in the scatter plots, that are well outside of the 99.9\% confidence intervals that would be expected if the data had been normally distributed. In file 335 (C, representative of the non-FOI files), no such extreme events are found in the residuals. The residual data points are marked as either blue if the residual is within the 99.9\% probability area, or red if the residual is outside this area. The ellipses in black and yellow show the 99\% and 99.9\% probability areas, respectively. These areas are the region which, under the null hypothesis of normally distributed data, contain 99\% and 99.9\% of all data points. The orientation of the residual extreme events agrees overall well with the dominating incoming waves direction. The direction of the wave field, found by determining the horizontal direction of maximal wave acceleration intensity in the lowpass-filtered wave field IMU data, is shown as the IMU wave direction lines. The arrow shows the estimated direction of the geographic north, computed using the magnetic decline using the World Magnetic Model \cite{chulliat2020us}. The direction of the incoming waves is also shown in Fig. \ref{fig:smap}, suggesting the waves are coming from the south in the two FOI. }
  \label{fig:scatterH}
\end{figure}

We present the residuals in the 2D horizontal plane in Fig. \ref{fig:scatterH} for the same three files, including the two FOIs, plus F335 where no extreme residual values are detected. Each time series is twenty minutes long with a measuring frequency of 10 Hz. Hence, the expected amount of points outside the 99.9 \% probability region is 12 if the residual data are normally distributed. This is not the case in the two FOIs, F291 and F312, where there are 108 and 67 extreme residual events respectively.  This is shown in sub-figures (A) and (B) in Fig. \ref{fig:scatterH}, where the purple ellipse shows the 99 \% probability region and the yellow one encompasses the 99.9\% probability region. By contrast, this is approximately the case for the reference file F335, where there are 10 extreme residual events. These extreme residuals are, in both FOI cases, generally aligned within a single consistent direction (different between F291 and F312), which corresponds generally with the dominating incoming wave direction obtained from the WAM-4km model data, as also visible in Fig. \ref{fig:smap} further down. We note that minor orientation differences between the incoming wave field and the collisions can be explained by, e.g., the specific shapes and relative positions between adjacent ice floes.

\begin{figure}
  \centering
  \includegraphics[width=0.99\textwidth]{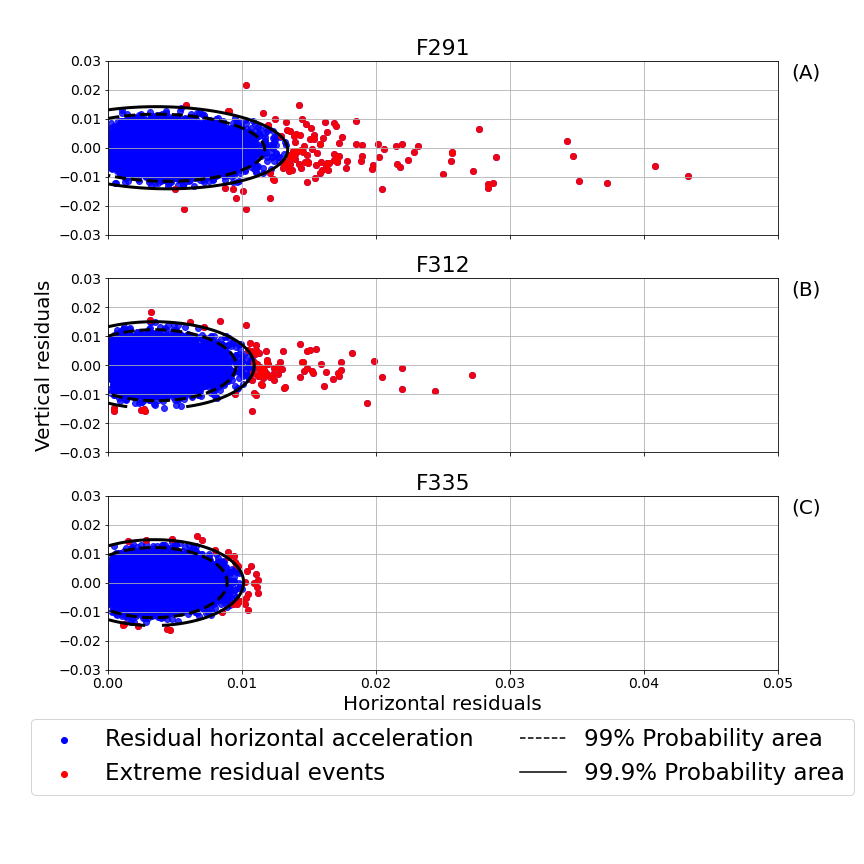}
  \caption{Illustration of the horizontal residuals vs the vertical residuals in three different time series. Clear extreme events are observed in the FOIs (files F291 (A) and F312 (B) where we discovered residual extreme events), similarly to Fig. \ref{fig:scatterH}. By contrast, no residual extreme events are found in file 335 (C). The ellipses show the 99\% and 99.9\% probability areas under the hypothesis that the residuals are normally distributed, similarly to Fig \ref{fig:scatterH}. Strikingly, the extreme events correspond to large values of the horizontal, but not the vertical, residuals. This corresponds well to what would be expected from the collision between adjacent floes.}
  \label{fig:scatterV}
\end{figure}

We find, upon closer investigation, that the extreme residual values in the FOIs are caused by the signal in the horizontal, rather than vertical, direction. In Fig. \ref{fig:scatterV} we show the scatter plot of the vertical acceleration versus the magnitude of the horizontal acceleration. The horizontal acceleration is defined as
\begin{equation}
    a_h = \sqrt{a_x^2 + a_y^2}. \label{eq:ahdef}
\end{equation}
The predominantly horizontal component of the extreme residual values shown in Fig. \ref{fig:scatterV}, is the expected pattern for ice floes colliding with their neighbours.

\begin{figure}
    \centering
    \includegraphics[width=0.99\textwidth]{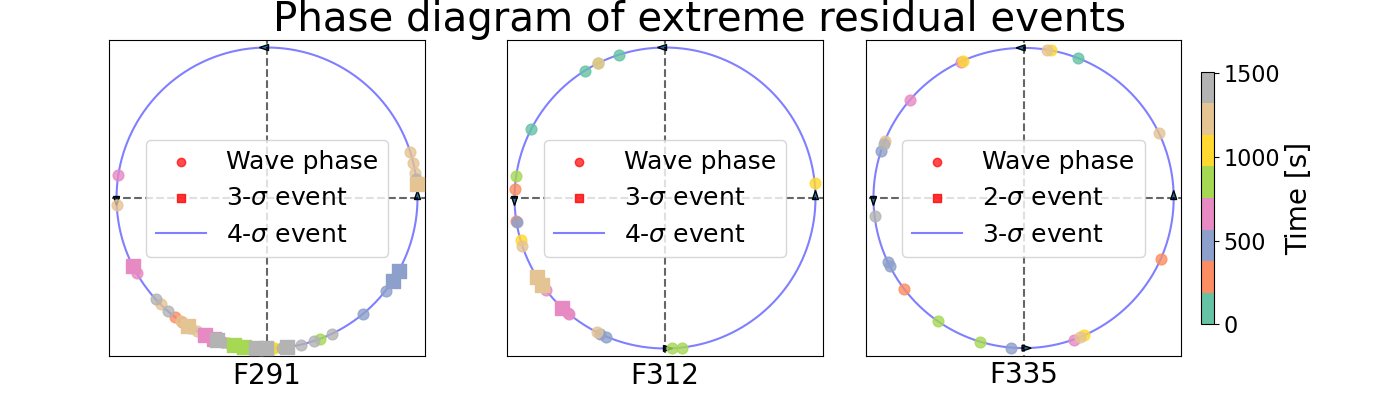}
    \caption{Phase diagram of the extreme residual values detected the two FOIs and the control file F335 where no collision like events were detected, relative to the incoming wave phase. The phase of the wave field is found using a Hilbert transform. Then, each extreme residual event (meaning it is outside the 99.9\% probability region) is plotted at its respective phases. The upper half of the circle is the wave crest, while the bottom half is the wave trough. In the two FOI, we see signs of  phase clustering, indicating that the occurrence of the extreme events we detect is strongly correlated to specific portions of the incoming wave phase motion. The extreme events in F335 are both less extreme (no $4\sigma$-events were observed) and less clustered in the phase diagram.
    }
    \label{fig:phase}
\end{figure}

The extreme values in the residuals are likely to be caused by floe-floe collisions induced by the incoming waves. This is supported both by the general alignment of the residual acceleration extreme events in the horizontal plane with the dominating incoming wave direction, and the absence of extreme residuals in the vertical direction. In order to further cross-check this hypothesis, we look at the correlation between the instantaneous wave phase obtained by the Hilbert transform of the double integrated wave elevation $\eta(t)$, and the occurrence of extreme events in the acceleration residuals. If the extreme residual events are wave-induced collisions, we would expect the majority of the extreme events to happen at the the same point during the wave phase, as that point in the wave phase corresponds to the convergence of the colliding ice floes . As visible in Fig. \ref{fig:phase}, the extreme events are not uniformly distributed, with the majority of collisions in F291 happening in the third and fourth quadrants of the phase diagram and F312 in the second and third quadrants, and hence the extreme events follow the phase of the wave field.  The extreme events of the control file, F335, are both not as extreme as the ones in the two FOIs, and do not exhibit any clear clustering in the data.

Following the evidence presented above, it appears most likely that the extreme residual events recorded in the FOIs correspond to wave-induced collisions between adjacent floes.

\subsection{Waves and sea ice conditions when collisions happen}

\begin{figure}
    \centering
    \includegraphics[width=0.9\textwidth]{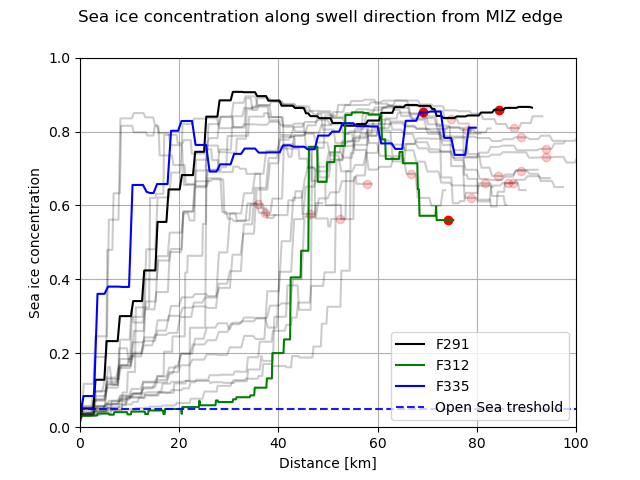}
    \caption{Sea ice concentration (SIC) encountered by waves propagating into the MIZ following the direction of the swell. The SIC is read from the ice model, while the closest available swell direction is read from the wave model. The blue dashed line shows the threshold corresponding to a SIC of 5\%. The red dots indicate the buoy position along the swell direction relative to the edge of the MIZ. All SIC profiles are aligned to start right at the start of the MIZ (according to the 5\% SIC threshold criterion) at the 0 km distance. The two FOIs and the control file F335 are highlighted in a dark color. The buoy is, in most files, more than 60 km from the edge of the MIZ alongside the propagating swell. }
    \label{fig:iceplot}
\end{figure}

A key aspect of our observations is that they are obtained deep into the MIZ. This is made visible by Fig. \ref{fig:iceplot}: the buoy recovered is typically around 80 km further in the ice compared with the location of the 5\% SIC limit (open sea threshold of the figure), and 40 to 60km further in compared to the 25\% SIC threshold. While the SIC at the location of the buoy fluctuates in time according to the model, which may be attributed to the effect of winds and currents on the level of sea ice packing, the SIC is typically over 80\% at the location of the buoy and already around 80\% for approximately 55 km (respectively 35 km) in the MIZ area towards the open ocean in front the buoy, for the two FOIs, respectively.

\begin{figure}
  \centering
  \includegraphics[width=0.8\textwidth]{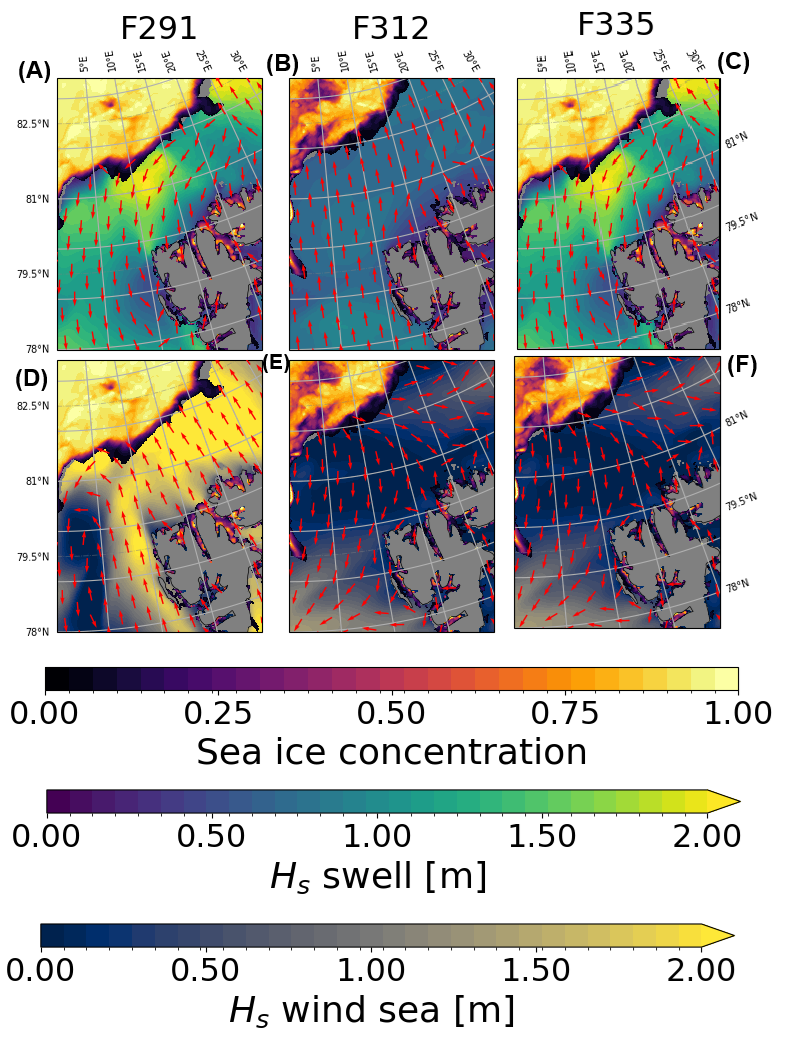}
  \caption{Situation map showing the sea state for the two FOIs at the time when the collisions occurred. The sea ice concentration is shown in all four figures. The top row (figures (A),(B) and (C)) shows the model swell partition significant wave height and direction at the time of F291 (left), F312 (middle) and F335 (right), respectively, while the bottom row (figures (D), (E) and (F)) similarly shows the model wind sea partition significant wave height and direction for the two FOIs and control file. The position of the buoy at the time corresponding to each file is marked with a light blue star.}
  \label{fig:smap}
\end{figure}

A synoptic view of the wave conditions in the area is presented in Fig. \ref{fig:smap}. In both cases, there is significant wave energy incoming in the MIZ. In the case of F291, the incoming wave energy comes mostly from the locally generated wind sea partition according to the wave model. However, there is significant fetch available, so this results in a $H_s$ of up to around 2m and the existence of low frequency components in the wave spectra generated, as visible in Fig. \ref{fig:PSD}. In the case of F312, the incoming wave energy comes mostly from the long traveling swell partition according to the wave model, which is in good agreement with the observation of waves of lower peak frequency, as visible in Fig. \ref{fig:PSD}. In addition, the $H_s$ is typically quite a bit lower, down to typically 1 m outside of the MIZ for the swell heading to the buoy, corresponding well to the lower energy levels observed in the spectrum from F312. This confirms that, in both cases, wave energy incoming from the South (respectively South-South-East) is expected at the location of the buoys, in good agreement with the general horizontal direction for the collisions observed in Fig. \ref{fig:scatterH}.

\begin{figure}
  \centering
  \includegraphics[width=0.9\textwidth]{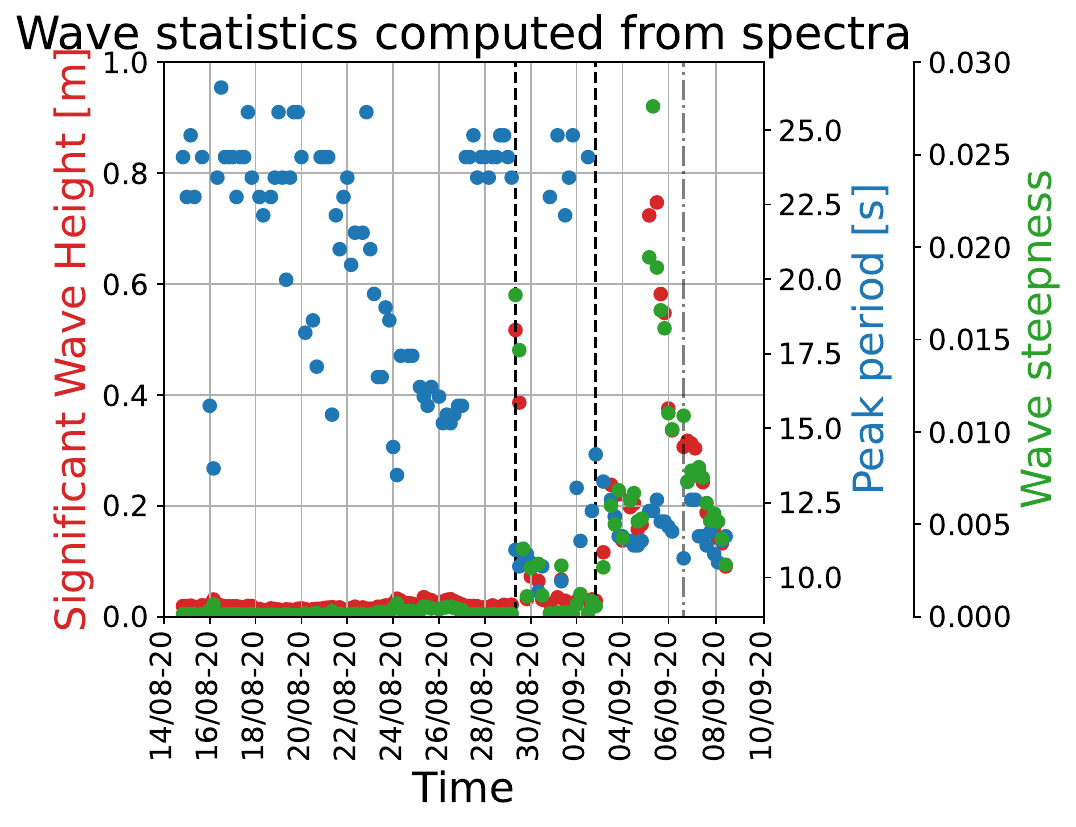}
  \caption{Significant wave height (red), peak period (blue) and wave steepness $a_k$ (green) obtained from processing of the raw data files obtained from the SD card of the recovered buoy as a function of time. The black dashed lines indicate the times for which we detect collisions, while the gray dashed and dotted line is the control file F335, where no collisions were found.}
  \label{fig:Wavestats}
\end{figure}

In Fig. \ref{fig:Wavestats}, the peak period and significant wave height for each time series is shown. A significant number of files in the early part of the time series contain very little wave motion, and as a consequence the recorded time series are dominated by sensor noise and $T_p$ saturates to very long periods due to the low frequency noise implied by the double time integration (similar to what is described in, e.g., \citet{rabault2022openmetbuoy}). The corresponding samples are indicated with a transparent color in Fig. \ref{fig:Wavestats}.

\section{Discussion}

The statistical Chi-squared test we implement discovers clear extreme residual acceleration events in the raw time series recovered from our stranded ice buoy. In particular, the null hypothesis of normally distributed residuals was rejected in four files with a confidence of 99\%. Upon closer analysis, in two of these four files, we find extreme acceleration events outside the 99.9\% probability area in the horizontal plane, that are both generally aligned with the incoming wave direction, and are phase locked with the wave signal recorded by the buoy. Therefore, we have compelling evidence suggesting that these acceleration events correspond to wave-induced floe-floe collisions.

While we have convincing evidence that our records contain wave-induced floe-floe collisions, it is difficult to be more specific about the detailed mechanics of these collisions. While the motion of the sea ice under the influence of waves is mostly happening in the vertical direction, some residual horizontal motion can also take place. Therefore, if neighboring ice floes are close enough, it is possible that this residual horizontal motion can lead to floe-floe collisions, as recently discussed in \citet{rabault24collisions}. While one could expect that adjacent floes with similar shapes move in (near) synchronization under the influence of monochromatic waves, conditions in the field can lead to the existence of significant relative motion between floes. In particular, real world wave fields are not monochromatic and individual waves have a spread in both their individual amplitude and wavelength. This naturally leads to different forcing between consecutive waves, which can result in different motion between adjacent floes at any given time, and, hence, collisions. In addition, ice floe sizes typically follow a relatively broad floe size distribution \citep{toyota2006characteristics}. This means that adjacent floes can have different responses to the same incoming waves, which, in turn, can also be a source of collisions. However, more work will be needed to investigate the detailed underlying mechanics, possibly by combining buoy observations with other more exhaustive techniques.

Extreme acceleration events acting on ice floes in the MIZ have also been investigated by \cite{laderach2011seismic}. In that study, the authors used seismometers to investigate earthquakes in the area surrounding the Gakkel ridge north of Svalbard. While their study explicitly only considers the seismological activity, the authors note they discovered several icequakes in the data, which they postulate could stem from ice dynamical phenomena such as cracking or floe-floe collisions. The ice quakes are, similarly to the extreme events discussed in this article, mainly aligned in the horizontal plane, which clearly separates them from the vertical movements associated with earthquakes. However, the presented data in \citet[figure 6]{laderach2011seismic}, does not seem to be caused by the same mechanisms governing the extreme events we present in e.g. Fig. \ref{fig:detectedcollisions}. The icequakes discovered by \citet{laderach2011seismic} are shorter in duration, at $\approx$ 10 seconds (subfigure 6 (B) of \citet{laderach2011seismic}), compared to the periodic spikes in Fig. \ref{fig:detectedcollisions} which occur routinely over a time period of over a minute. Furthermore, the icequakes exhibit a significantly higher frequency than what is seen in Fig. \ref{fig:detectedcollisions}, suggesting other mechanisms than wave in ice activity caused the ice quakes presented in \citet{laderach2011seismic}. 

Another interesting finding is that the collisions we detect do not seem to be strongly correlated with the significant wave height. More specifically, we find, surprisingly, no clear relationship between $H_s$ and the existence of collisions in the data. A possible explanation is that our collision detection criteria is too strict, and that collisions are happening in high $H_s$ situations, but the sigma-filtering criterion we use is not sensitive enough to detect these and only detects the strongest and most obvious collisions. More work, and more advanced statistical analysis, may be needed to test this hypothesis further. However, this may also be explained by $H_s$ not being a predictor for the presence of collisions. This explanation is supported by our observation that the collisions we detect happen when a sudden, drastic change in the local conditions takes place. More specifically, the collisions we detect here happen when a sudden jump in $H_s$ and peak wave period take place, as shown in Fig. \ref{fig:Wavestats}. This may indicate that collisions happen mostly when the sea state is "out of balance" with the incoming wave conditions, i.e. when sea ice needs to "adapt" to sudden changes in the forcing acting on it.

Therefore, while the waves are a key ingredient in the local mechanisms at the origin of the collisions we observe, as our data indicate that waves are the local driver for individual collision events, wave activity does not seem to be enough to induce floe-floe collisions. Following this observation, there must be additional factors, either local or influencing the ice sheet at a larger scale, that contribute to creating conditions that are favorable for wave-induced floe-floe collisions. We observe no particular periodicity in the occurrence of collisions in our data, which indicates that tides and tidal currents by themselves are not a likely large scale mechanism to produce the conditions favorable to collisions. Moreover, we have investigated (not reproduced here for brevity) the relation between the observation of collisions in the FOIs and the sea ice drift pattern recorded by the GPS at 1Hz during the 20 minutes wave measurement intervals. There also, we were not able to find a particular pattern or relation between the sea ice drift velocity at the location of the buoy by itself, and the presence or absence of collisions. Therefore, while sea ice drift may be an enabling factor to create conditions that allow for wave-driven collisions, the local sea ice drift velocity of an individual floe does not appear to explain for the occurrence of collision events. Similarly, we investigated the wind conditions over the whole MIZ area around the buoy (not reproduced here for brevity), and we found that the wind conditions are very different between the two FOIs (in the case of F291 the typical wind speeds and direction in the MIZ area around the buoy are 10 m/s towards the ice, while for F312 we observe typically down to around 5m/s towards the open water), so that there is no simple link between wind and the occurrence of collisions either. Interestingly, recent works by our group \citep{rabault24collisions} indicate that more complex large scale ice sheet properties resulting from a combination of forcings, such as the general patterns of sea ice convergence and divergence, may play a role in modulating the wave in ice attenuation damping and possibly the presence of collisions in some circumstances. However, more work and the gathering of larger, statistically representative datasets, will be needed to further investigate this hypothesis and to provide data-based evidence on what ice sheet properties create favorable conditions for floe collisions under the influence of waves.

The absence of more buoy measurements also means we are not able discuss the attenuation of the wave field in details. Furthermore, we are not able to discuss the prevalence of collisions in details either, as while we have found evidence of floe-floe collisions in two files,  this is not a sufficiently large and well-sampled dataset to extrapolate our results. In the best case scenario, the data might give an estimate of the collision rate of an ice floe of the same size (which is unknown) as the floe at which the buoy was deployed. But more critically, the data we have indicate the collisions to be wave-induced. Hence, when over half the measurements were taken during conditions with little to no wave activity (see figure \ref{fig:Wavestats}), any discussion on the frequency of collisions and their possible attenuating effect  is challenging from these data alone. However, we want to stress that our present observations fit well with previous results. \citet{loken2022experiments} ran experiments where two ice floes were rammed into each other using winches, and studied the generation of TKE from the collisions and found that $36.9\% \pm 23.7\%$ of the input energy was dissipated through turbulence. Furthermore \citet{herman2019wave} found the wave attenuation induced by colliding ice floes to be proportional to the square of the orbital velocity underneath. A model for the attenuation of ice floes was published by \citet{shen1998wave}, in which a theoretical model for the attenuation by collisions between pancake floes is derived. Our results can not be directly compared to their model, as we are not able to infer the attenuation rate of the wave field and the floe size and distribution in the area. Yet, we still observe that \citet{shen1998wave} found a clear wave amplitude dependency on the collision frequency, with all three model runs predicting low collision frequencies $f < < 1$ Hz when the wave amplitude is low enough. We do, however, still note that the model of \cite{shen1998wave} assumes pancake floes at a given size. 

The present study owes a lot to luck and serendipity. Both the fact that the instrument v-2018 contains a SD card with a full copy of the time series (which is due to the fact that the microcontrollers available back then were not able to perform the signal processing needed, so that an additional RaspberryPi had to be added for processing the time series), and the fact that the instrument recovered got stranded, was found, and that the SD card was not damaged, are a strike of luck. However, it would be easily possible, based on the experience obtained from the present dataset, to adapt existing open source instruments firmware, such as the OMB-v2021, to i) save raw time series on an embedded SD card in case the instrument is found again, ii) compute in-situ a number of collision statistics, such as the ones used here, and append these to the wave spectra that are being transmitted. Since a simple metrics of collisions, such as the number and relative strength of extreme acceleration events and their correlation with the wave phase, or a deviation in the distribution from the expected distribution of wave-induced acceleration measurements (as implemented by e.g. \citet{brown2017hydrodynamic}),  would be only a few scalars, this would not add prohibitive transmission costs nor energy consumption to the design. Moreover, all the data needed to perform such an analysis are already available on-board. Adding such metrics to wave in ice buoys transmissions would allow to perform large scale, statistically representative analysis of the correlation between collision-like proxy metrics, changes in the sea ice and wave in ice conditions, and the wave in ice attenuation rate.

We also observe that the present study reaches several conclusions that seem to converge with the findings of \citet{rabault24collisions}. There, the authors observe strong modulation in the wave in ice $H_s$, and proceed by elimination to suggest that the switching on and off of floe-floe interactions by large scale sea ice convergence and divergence patterns may be a possible mechanism. The present study confirms that the existence of collisions is a realistic mechanism, as discussed in the study of \citet{rabault24collisions}.

\section{Conclusion}

We recovered raw data time series of wave-in-ice motion from a buoy that got stranded on the northern Icelandic coast. Statistically significant extreme residual acceleration events, i.e., spikes in the acceleration signal once the incoming wave motion is subtracted, are detected with a statistical confidence of 99\% in four of the 20 minute long raw data files. Of these four, two are then found to have clear signs of collisions in the raw data. These manifest as extreme residual acceleration events corresponding to spikes in the horizontal acceleration measured by the buoy. The direction of these spikes is generally aligned with the dominating wave direction incoming from the open ocean that is predicted by ocean models, and corresponds well to the local wave direction observed in the horizontal in-situ acceleration time series. Moreover, the occurrence of the acceleration events is phase locked to the incoming wave signal.

Based on this evidence, we conclude that the data we recovered contain wave-induced floe-floe collisions between the floe on which the buoy was sitting, and neighboring floes. This is the first time, to our knowledge, that such clear floe-floe collision evidence are provided deep into the Marginal Ice Zone (MIZ). This proves that collisions between adjacent ice floes can happen deep in the MIZ, and that confirms that floe-floe collisions are a relevant mechanism that can contribute significantly to the wave energy dissipation and the associated increased water turbulence levels, including in the deep MIZ.
Based on these evidence, we conclude that the data we recovered contain wave-induced floe-floe collisions between the floe on which the buoy was sitting, and neighboring floes. This is the first time, to our knowledge, that such clear floe-floe collision evidences are provided deep into the Marginal Ice Zone (MIZ). This proves that collisions between adjacent ice floes can happen deep in the MIZ, and that confirms that floe-floe collisions are a relevant mechanism that can contribute significantly to the wave energy dissipation and the associated increased water turbulence levels, including in the deep MIZ.

Interestingly, the occurrence of collisions does not seem to be determined by the value of the significant wave height or the peak period. Instead, the collision events that we discover happen at times when a sudden change in the wave in ice conditions is measured by the instrument. However, we observe that signals that visually resemble collision-like events are present also in files where our statistics-based test does not detect extreme events with a 99\% confidence. This may correspond to weaker collisions, though more investigation would be needed to determine if this is truly the case. More data will be necessary to establish statistically representative evidence of what large scale ice sheet properties and weather conditions create favorable conditions for wave-driven floe-floe collisions.

The methodology we used to detect the extreme acceleration events can be simply applied on the raw time series of the acceleration measurements, and its result can be summarized into a few key parameters. Therefore, we plan to work in the future on implementing embedded routines that perform similar processing on our buoys, such as the OpenMetBuoy, and we plan to transmit these metrics back over iridium together with the wave spectra. We hope that this will allow, over time, to build a statistically representative dataset that will i) provide a better understanding of collision occurrence in the MIZ, ii) make it possible to compare the occurrence of collisions to changes in the wave in ice damping coefficient, iii) elucidate what conditions lead to the occurrence of floe-floe collisions.

\section*{Acknowledgements}

J.R. and Ø.B. gratefully acknowledge the Nansen Legacy Project, funded by the Norwegian Research Council to the Norwegian Meteorological Institute (NFR grant AeN, NFR-276730). J.R., A.J., acknowledge funding from the Dynamics Of Floating Ice project funded by the Norwegian Research Council to the University of Oslo (NFR grant DOFI, NFR-280625). Y.K. gratefully acknowledges funding and expedition support from Blodgett-Hall Polar Presence LLC and Lundin Energy Norge.

\section*{Appendix A: Data and code availability}

All the raw data recovered on the stranded buoy are made available at \url{https://github.com/larswd/MIZ\_Floe\_collisions\_archive} [will be made available upon publication in the peer reviewed literature]. We also provide the scripts used to analyze the data and generate the figures used in this paper at the following location: \url{https://github.com/larswd/MIZ\_Floe\_collisions\_archive} [will be made available upon publication in the peer reviewed literature].

\bibliographystyle{unsrtnat}
\bibliography{bibliography.bib}

\end{document}